\pdfoutput = 1
\documentclass[%
 reprint,
 prd,
 amsmath,amssymb,
 aps,
 superscriptaddress,
 floatfix
]{revtex4-1}

\usepackage{dcolumn}
\usepackage{bm}
\usepackage{ulem}
\usepackage{multirow}
\usepackage{placeins}
\usepackage{algorithmic}
\usepackage{color}
\usepackage{hyperref}

\begin{document}

\title{Hawking radiation of anyons}
\author{Vishnulal C}\thanks{lal15@alumni.iisertvm.ac.in}
\affiliation{School of Physics, Indian Institute of Science Education and Research Thiruvananthapuram, Maruthamala PO, Vithura, Thiruvananthapuram 695551, Kerala, India}
\author{Soumen Basak}\thanks{sbasak@iisertvm.ac.in}
\affiliation{School of Physics, Indian Institute of Science Education and Research Thiruvananthapuram, Maruthamala PO, Vithura, Thiruvananthapuram 695551, Kerala, India}
\author{Saurya Das}\thanks{saurya.das@uleth.ca}
\affiliation{Theoretical Physics Group and Quantum Alberta, Department of Physics and Astronomy, University of Lethbridge 4401 University Drive, Lethbridge, Alberta T1K 3M4, Canada}

\date{\today}

\begin{abstract}
We derive the Hawking radiation spectrum of anyons, namely particles in $(2+1)-$dimension obeying fractional statistics, from a Bañados, Teitelboim, and Zanelli (BTZ) black hole, in the tunneling formalism. We examine ways of measuring the spectrum in experimentally realizable systems in the laboratory. 
\end{abstract}

\maketitle

\setlength{\columnsep}{0.5cm}

\section{Introduction}
According to the classical theory of Gravity, formulated by Albert Einstein, a black hole is a region of spacetime from which gravity prevents everything, including light, from escaping. It is quite remarkable that black holes turn ``grey" and radiate energy in the form of Hawking radiation \cite{article6}, when quantum mechanics is brought into the picture. Hawking radiation, being thermal in nature, contains very little information, and since a large amount of information may have entered the black hole during its formation phase, it may be lost forever when the black hole evaporates completely. This is the so-called information loss problem, whose resolution is still-being sought, despite a number of interesting proposals \cite{article5, article7, article8, article30, article31, article29}. Hence, it is important to examine Hawking radiation phenomena for a diverse set of spacetimes and a variety of particles. At the very least, this would help in a better understanding of the problem itself. 
It has been shown that in $(3+1)-$ dimension, bosons and fermions Hawking radiate from black holes at the {\it same} Hawking temperature of the black hole, with the respective Planck and Fermi distributions. Contrary to this, in  $(2+1)-$dimension, there can exist particles, known as {\it anyons}, which are neither bosons, nor fermions, but follow fractional statistics \cite{article11,article12}.
Anyons are interesting entities to study in their own right. Furthermore, there has been recent experiments which show  strong evidence in favour of their existence \cite{article35}. They may also be practically useful in a variety of systems such as quantum computation \cite{article36,article37}. 
The obvious question that arise in this context is whether there exist Hawking radiation of anyons as well. Existence of Hawking radiation of anyons from black holes will not only strengthen the 
Hawking radiation results, but may also shed new light on the information loss problem, as well as provide a new avenue of observing Hawking radiation in the laboratory \cite{article9}. Hence the primary focus of this article is to take a closer look at these issues, which to the best of our knowledge, is the first study of anyonic Hawking radiation. 
In the next section, we review the important properties of anyons that are relevant to this work. This is followed by, in Section \ref{hrsection}, a brief review of the Hawking decay rate from a black hole horizon in the tunneling approach first proposed by Parikh and Wilczek \cite{article2} for bosons and fermions in $(3+1)-$dimension. Hawking radiation using a complex path approach in different coordinates was also studied in 
\cite{article43, article44} and anyon like excitations
in the context of Bañados, Teitelboim, and Zanelli (BTZ) black holes in \cite{Luo:2017ksc}.
In Section \ref{hranyonsection}, we describe in detail an extension of this formalism to anyons in the context of a BTZ black hole. In Section \ref{exptsection}, we examine potential ways of observing this radiation in tabletop experiments. 
Finally, we summarize our results and conclude in Section \ref{conclusionsection}. 
\section{Particles with intermediate statistics}
In the $(3+1)$-spacetime dimension that we live in, particles are either bosons or fermions, with intrinsic spin 
of integer or half-odd integer (in units of $\hbar$). These particles are described by wave functions which are either symmetric or anti-symmetric under the exchange of two particles. Contrary to this, in $(2+1)$-spacetime dimension, a continuous range of statistics is available \cite{article12}.
Consider two identical particles in $(2+1)$-dimension. Let $\psi(r)$ be the wave function of the two particle system,
subject to the condition that $\psi(r) \neq 0$, if $r >a$ (the so-called `hard-core condition'), where $\vec r_1$ and $\vec r_2$ are the positions of the two particles and the relative position vector $\vec r \equiv \vec r_{1}-\vec r_{2}$. So, the configuration space of the particles is the two-dimensional $(x,y)$-plane with a disc of radius $\mathbf{a}$ removed. 
Given these coordinates, we can define a complex coordinate $z=x + i y$, and make a transformation $z\rightarrow z\,e^{2\pi i}$, which effectively brings a particle back to its starting point, the wave function must also remain invariant, {\it but} up to a phase, i.e. 
\begin{eqnarray}
   \psi(z e^{i2\pi} , z^{*} e^{-i2\pi}) = e^{i 2 \pi \alpha} \psi(z,z^{*}) ~,
    \label{Eq1} 
\end{eqnarray}
for some real parameter $\alpha$. %
Similarly, one may interchange the two particles, i.e. transform $z \rightarrow z e^{i \pi}$, to obtain,
\begin{eqnarray}
    \psi(z e^{i  \pi} , z^{*} e^{- i \pi} ) = e^{i \pi \alpha} \psi(z,z^{*})~.
    \label{Eq2}
\end{eqnarray}
In Eq.(\ref{Eq2}) the value of $\alpha$ equal to $0$ and $1$ corresponding to bosons and fermions. However, in $(2+1)-$dimension, any real value of $\alpha$ in between $0$ and $1$ is also allowed. To understand this particular point, let's consider a system of two identical particles in $(3+1)-$dimension. We need to do two consecutive interchanges of the location of the particles to go back to the original configuration. All such trajectories are topologically equivalent. However, in (2+1) dimension, after doing one interchange, we need to do one more winding of one particle around the other to come back to the initial configuration. Unlike to the case in $(3+1)-$dimension, these two trajectories are distinct in nature. So, it can be associated with two different topological phases which can take any value. It clearly indicates to the fact that, in two spatial dimension, a particle may posses statistics which are different from the standard Bose-Einstein and Fermi-Dirac statistics and are called fractional statistics. The particles that follow fractional statistics are called anyons. The objective of our work is to describe the Hawking radiation of these particles.
\section{Hawking radiation from tunneling for bosons and fermions}
\label{hrsection}
Hawking radiation \cite{article2} from a black hole horizon can be interpreted in the following way: there is copious pair production of particles and anti-particles from vacuum just inside the horizon. 
The anti-particle travels backwards in time inside the horizon, while the particle tunnels out quantum mechanically and is manifested as Hawking radiation. An equivalent picture exists in which the pair production occurs just outside the horizon, with the negative energy particle tunneling inside the horizon and the one with positive energy gives rise to Hawking radiation.
A rigorous calculation of the tunneling rate indeed reproduces the correct radiation rate, derived independently by other methods, and lends further credence to the above picture. 
To estimate the tunneling probability, we compute the imaginary part of the action of the particle over classically forbidden region \cite{article2},
\begin{eqnarray}
    Im  \,\,  S &=& I m  \int_{r_{in}}^{r_{out}}  dr \, p_{r}\,, \nonumber\\
                        &=&  I m  \int_{M}^{M-\omega} \int_{r_{in}}^{r_{out}} dH \, \frac{dr}{\dot r} \,,  
    \label{eqbeg}                    
\end{eqnarray}
where $H = M - \omega^{'}$ and the initial value of radius is chosen just inside the event horizon, $r_{in} = 2M - \epsilon$. Due to loss of energy, the radius of the event horizon will reduce to, $r_{h} = 2M - \omega$. Therefore, the point outside will be at, $r_{out}=2(M-\omega) + \epsilon$. 

The integration in Eq.(\ref{eqbeg}) is obtained using the equation of motion of radial null geodesics to express the $I m  \,\,  S$ in terms of $M$ and $\omega$,
\begin{eqnarray}
    Im \,\, S = 4 \pi \omega \left(M - \frac{\omega}{2}\right)~.
    \label{eq3}
\end{eqnarray}
The corresponding amplitude of the tunneling process can be written in a straightforward manner as,
\begin{eqnarray}
    \Gamma \displaystyle \sim e^{-2 Im \hspace{5pt} S} = e^{-8 \pi \omega \left(M - \displaystyle{\frac{\omega}{2}}\right)}    
     = p_{\omega} ~.
     \label{eq10}
\end{eqnarray}
The above tunneling amplitude also can be interpreted as the relative probability of creating a particle-antiparticle pair just outside the horizon \cite{article41}. Eq.(\ref{eq10}) is valid for both bosons and fermions. 
When a pair production happens just outside the horizon with the relative probability $\rm{p_{\omega}}$, particle with energy $\omega$ escapes from the horizon and antiparticle with energy $-\omega$ goes inward towards the singularity.
\noindent The absolute probability of creating a pair of particles in a particular mode $\omega$ is calculated as follows:
\begin{eqnarray}
    \rm{P_{\omega} \, = \, C_{\omega}\,p_{\omega}}~,
\end{eqnarray}
where $\rm{C_{\omega}}$, which is the probability that no pairs are created in that particular mode. Since fermions obey the exclusion principle, only one particle-antiparticle pair can be created in each quantum state, the sum of the probabilities of no pair production and one pair production must add up to one,
\begin{eqnarray}
    \rm{C_{\omega} \, + \, C_{\omega}\,p_{\omega} \, = \, 1 }~.
\end{eqnarray}
On the other hand, the probability of creating one pair of particles in the energy mode $\omega$ is given by
\begin{eqnarray}
    \rm {P_{1\omega} \, = \, C_{\omega}\,p_{\omega}} \, = \, \frac{p_{\omega}}{1+p_{\omega}}~.
\end{eqnarray}
We know that there is an effective potential barrier exterior of the black hole ($ \rm{2M \,  < \, r \,  <  \, \infty}$) which causes a back-scattering. In the particle production calculation we only need to consider the fraction entering black hole horizon, which is represented by the transmission coefficient $\Gamma_{\omega}$ 
\cite{article16}.
Taking this fact into account we compute the probability of the one particle emission,
\begin{eqnarray}
    \rm{\Bar{P}_{1\omega} \, = \, P_{1\omega}\,\Gamma_{\omega} \, = \, \frac{p_{\omega} \, \Gamma_\omega}{1+p_{\omega}}}~,
    \label{eq14}
\end{eqnarray}
$\rm{\Bar{P}_{1\omega}}$ can also be interpreted as the mean number of particles
($\bar N_{\omega}$) emitted in a given mode. 
We substitute Eq.(\ref{eq10}) in Eq.(\ref{eq14}) 
to obtain $\bar N_{\omega}$ \cite{article13},
\begin{eqnarray}
   \rm{\bar N_{\omega} \, = \, \frac{\Gamma_{\omega}}{e^{8\pi m\omega} \, + \, 1}} ~.
   \label{eqfer}
\end{eqnarray}
The Eq.(\ref{eqfer}) is precisely the Fermi-Dirac distribution modified by the transmission coefficient $\Gamma_{\omega}$. Similar expression can be derived for bosons as well \cite{article13}
\begin{eqnarray}
    \rm{\bar N_{\omega} \, = \, \frac{\Gamma_\omega}{e^{8\pi M\omega} \, - \, 1}} ~. 
    \label{eq5}
\end{eqnarray}
These results are consistent with Hawking's celebrated work about particle emission from black hole and provides a nice physical picture in terms of tunneling through a potential barrier \cite{article6}.
\section{Hawking radiation from tunneling for particles with intermediate statistics}
\label{hranyonsection}
Following the procedure described in section \ref{hrsection}, we now derive the amplitude for the emission of anyons from BTZ black hole. Since anyons exist in $(2+1)$-dimensional spacetime only, we restrict ourselves to BTZ black hole. This is a solution of the Einstein field equation in (2+1)-dimensional spacetime and describes a rotating geometry with horizons \cite{article18,article15}. The action corresponding to this solution is expressed in terms of metric $g^{(3)}$ and Ricci scalar $R^{(3)}$ in $(2+1)$-dimension as follows \cite{article18,article3},
\begin{eqnarray}
    S = \int d^{3}x \sqrt{-g^{(3)}}\,( R^{(3)} + 2 \Lambda )~,
\end{eqnarray}
and the line element in polar coordinates is given by
\begin{eqnarray}
    ds^{2} = - f(r) dt^{2} + \frac{dr^{2}}{f(r)}    + r^{2} \left(d\theta - \frac{J}{2r^{2}}dt
    \right)^{2},  
    \label{eq18}
\end{eqnarray}
\begin{eqnarray}
     f(r) = - M + \Lambda r^{2} + \frac{J^{2}}{4 r^{2}}~.
     \label{eqf}
\end{eqnarray}
Here $M$ and $J$ are respectively the mass and angular momentum of the $3$-dimensional rotating black hole and 
$\Lambda$ is the cosmological constant. 

For the simplicity of the calculations, to start with, we restrict ourselves to the $J=0$ case. Generalization to $J\neq 0$ is straightforward and is discussed at the end of this section. We start with the transformation from the original coordinates $(t,r)$ to the Painlev\'e coordinates $(t_p,r)$,
\begin{eqnarray}
    dt = dt_{p} - \frac{1}{f(r)} \sqrt{1 - f(r)} \, dr~.
\end{eqnarray}
In terms of new coordinates and via a dimensional reduction, the line element in Eq.(\ref{eq18}) takes the following form,
\begin{eqnarray}
ds^{2} = -f(r)\,dt_{p}^{2} + 2 \sqrt{1 - f(r)} dt_{p} dr + dr^{2}~.
\label{btzpainleve}
\end{eqnarray}
Next, consider the Lagrangian for a 
massive particle in the above background, 
\begin{eqnarray}
    L &=& \frac{m}{2} g_{\mu \nu} \frac{dx^{\mu}}{d \tau} \frac{dx^{\nu}}{d \tau} , \nonumber\\
     &=& -\frac{1}{2} m f(r) (\Dot{t_{p}})^{2} + m \sqrt{1 - f(r)}\, \Dot{t_{p}} \Dot{r} + \frac{m}{2} (\Dot{r})^{2}~,
\end{eqnarray}
where in the last step we substituted the metric components from Eq.(\ref{btzpainleve}). Since $t_p$ is a cyclic coordinate, the conjugate momentum is conserved,
\begin{eqnarray}
\frac{\partial L}{\partial \dot t_p} =  -m f \Dot{t_{p}}  + m\sqrt{1 -f} \,\Dot{r} = -\omega = \mathrm{constant} ~.
    \label{eq23}
\end{eqnarray}
The negative sign on the right hand side of the Eq.(\ref{eq23}) represents the positive energy of the tunneling particle. 
Exploiting the fact that massive particles travel along time-like trajectories, we derive from Eq.(\ref{eq23}) the expressions of the derivatives of Painlev\'e  coordinates with respect to proper time of the particle under consideration,
\begin{equation}
    \Dot{t_{p}} = \pm \frac{1}{m f(r)} \sqrt{(1-f(r))(\omega^{2} - m^{2} f(r))} + \frac{\omega}{m f(r)}~,
\end{equation}
\begin{eqnarray}
    \Dot{r} = \pm \frac{1}{m} \sqrt{\omega^{2} - m^{2} f(r) } ~.
\end{eqnarray}
Given the expressions of the derivatives of Painlev\'e  coordinates, we rewrite the imaginary part of the action Eq.(\ref{eqbeg}) as follows,
\begin{eqnarray}
    Im \,\, S = - \, Im \int_{r_{in}}^{r_{out}} \int_{m}^{\omega} \frac{d \omega^{'}}{\frac{dr}{dt_{p}}} dr~.
    \label{Eq28}
\end{eqnarray}
Following methodology described in \cite{article2}, we exchange the order of integration in Eq.(\ref{Eq28}) and do the integration first over the radial coordinate, 
\begin{equation}
      \int_{r_{in}}^{r_{out}} \frac{1}{\frac{dr}{dt_{p}}}  dr= \int_{r_{in}}^{r_{out}} \frac{ \sqrt{(1-f)(\omega^{2}-m^{2}f)} + \omega}{f\sqrt{\omega^{2} - m^{2}f}} dr~.
     \label{eq29}
\end{equation}
Using the Taylor expansion of $f(r)$ around the radius of the horizon, we rewrite the right hand side of Eq.(\ref{eq29}),
\begin{equation*}
      \int_{r_{in}}^{r_{out}} \frac{1}{\frac{dr}{dt_{p}}}  dr = \int_{r_{in}}^{r_{out}} G(r) dr ~,
\end{equation*}
where the function $G(r)$ is given by
\begin{equation*}
    G(r) = \frac{ \sqrt{(1-f^{'}(r_{h}) (r - r_{h}))(\omega^{2}-m^{2}f^{'}(r_{h}) (r - r_{h}))} + \omega}{f^{'}(r_{h}) (r - r_{h})\sqrt{\omega^{2} - m^{2}f^{'}(r_{h}) (r - r_{h})}} dr~.
\end{equation*}
The final expression of this integral is obtained using Cauchy's residue theorem,
\begin{eqnarray}
      \int_{r_{in}}^{r_{out}} \frac{1}{\frac{dr}{dt_{p}}}  dr =- \frac{2 \pi i}{f^{'}(r_{h}) } ~.
\end{eqnarray}
In terms of this integration, the imaginary part of the action simplifies to the following form,
\begin{eqnarray}
    Im \,\, S =   \pi \int_{m}^{\omega} \frac{2}{f^{'}(r_{h})} d\omega^{'} = \frac{\pi}{\sqrt{\Lambda M}}(\omega-m), 
\end{eqnarray}
where we have used the fact $r_{h}$ depends on $\omega$,
\begin{eqnarray}
r_{h}= r_{h} (M- \omega^{'})~.
\end{eqnarray}
The final expression of the tunneling rate for $J = 0$ is obtained using $Im \,\, S$,
\begin{eqnarray}
\Gamma \displaystyle \sim e^{-2 Im \,\,S} = e^{-\frac{2\pi}{\sqrt{\Lambda M}} (\omega-m)}~.
\label{eq37}
\end{eqnarray}
It is to be noted that we have ignored the back-reaction on the metric to get this results and hence, the results are valid when $m<<M$ and $\omega << M$.  From Eq.(\ref{eq37}) in the $m \rightarrow 0$ limit gives the tunneling amplitude for massless particle, 
\begin{eqnarray}
    \Gamma \displaystyle \sim e^{-2 Im \,\, S} &=& e^{-\frac{2\pi}{\sqrt{\Lambda M}} \omega}~.
    \label{eq38}
\end{eqnarray}
However, one can obtain the same expression starting from the condition for null geodesics and following the procedure applied in case of massive particles. 
It is to be noted that, to derive Eq.(\ref{eq37}), we have not explicitly used information about the statistics of the Hawking radiation particles. Therefore, it is valid not only for bosons and fermions, but also for anyons. The next step is to find the distribution function of anyons from the the expression of tunneling amplitude which we have already derived [Eq.(\ref{eq37})], which reduces to the standard Bose-Einstein and Fermi-Dirac distribution functions in appropriate limits. 
In order to find the expression of the distribution function, first we consider the following assumptions which hold for any particles in $(2+1)$-dimensions \cite{article1}.
\begin{itemize}
    \item The permutation of the coordinates of any two 
    particles in the multi-particle anyon wavefunction results in a phase being picked up by the wavefunction as follows (this is simply a restatement of the  Eq.(\ref{Eq2}) for anyons)
\end{itemize}
\begin{eqnarray}
    \Psi_{n}(. . ., q_{j}, . . .,q_{i},...) = f \, \Psi_{n} ( . . ., q_{i}, . . .,q_{j},...)~,
    \label{eq12}
\end{eqnarray}
\hspace{0.15in} where $f = e^{i \pi \alpha}$~.
\begin{itemize}
    \item Principle of detailed balance: 
    If $n_{1}$, $n_{2}$ are the mean occupation numbers for the states $1$ and $2$ respectively, then at equilibrium, the number of transitions from $1$ to $2$ is the same as from $2$ to $1$. Under this condition, the `enhancement factor' $F(n) \equiv P^{(n+1)}/P^{(n)}$, where $P^{(n)}$ is the probability of an $n$-anyon state, satisfies the following condition,
\begin{eqnarray}
    n_{1} F(n_{2}) e^{\frac{2 \pi}{\sqrt{\Lambda M}}(\omega_{1}-m)} = n_{2} F(n_{1}) e^{\frac{2 \pi}{\sqrt{\Lambda M}}(\omega_{2}-m)}~.
\end{eqnarray}
This condition implies that
\begin{eqnarray}
\frac{n}{F(n)} e^{\frac{2 \pi}{\sqrt{\Lambda M}}(\omega-m)} = \mathrm{constant}~.
\end{eqnarray}
\end{itemize}
The form of $F(n)$ can be read-off from
ref.\cite{article1}, which is $F(n)=\frac{4}{n+1} \left([\frac{n+1}{2}] \cos(\frac{\pi \alpha}{2})\right)^{2}$.
When expanded in a power series in $n$, this yields the following,
\begin{eqnarray}
    e^{\frac{2 \pi}{\sqrt{\Lambda M}}(\omega-m)} = \frac{1}{n} + a_{0}  + a_{1} n + a_{2} n^{2} + \dots ~.
    \label{occno1}
\end{eqnarray}
Inverting the above equation, we get the expression of occupation number as a function of $\omega$ and $m$,
\begin{eqnarray}
    n (\omega,m) &=& \frac{1}{g} + \frac{B}{{g^{2}}} + \frac{C}{g^{3}} + \frac{D}{g^{4}} + ...~, \nonumber\\
&\equiv & \frac{1}{g} + \sum_{k=3}^{\infty} \frac{\alpha_{k}}{g^{k}}~,
\label{Eq47}
\end{eqnarray}
where $g\equiv \displaystyle{e^{\frac{2 \pi}{\sqrt{\Lambda M}}(\omega-m)}} - a_{0}$ and the explicit form of the coefficients is taken from \cite{article1}.
Rewriting Eq.(\ref{Eq47}) in the form of a continued fraction, we get,
\begin{eqnarray}
    n(\omega,m) = \frac{1}{g - \frac{\alpha_{3}g}{g^{2}+\alpha_{3} - .....}}~.
\end{eqnarray}
Finally, using the expression of $g$ we find from the above equation, the generalized Hawking radiation formula in $(2+1)$-dimension,
\begin{eqnarray}
    n(\omega,m) = \frac{\Gamma_{\omega}}{e^{\frac{2\pi}{\sqrt{\Lambda M}} (\omega-m)}- a(\alpha)}~,
    \label{eq35}
\end{eqnarray}
where $\Gamma_{\omega}$ is the transmission coefficient. Eq.(\ref{eq35}) for $a(\alpha)=1$ and $a(\alpha)=-1$ corresponds to bosonic and fermionic Hawking radiation respectively. Any value of $a(\alpha)$ in between this range corresponds to anyons.\\
\noindent It is now quite straightforward to generalize the results to the $J\neq0$ case. Dimensional reduction of the full metric (Eq.(\ref{eq18})) gives the following 2-dimensional metric \cite{article20,article21},
\begin{eqnarray}
    ds^2 = -f(r) \, dt^2 + \frac{dr^2}{f(r)} , 
\end{eqnarray}
where
\begin{eqnarray}
     f(r) = -M + \Lambda r^2 + \frac{J^2}{4r^2} ~.
\end{eqnarray}
Following the procedure mentioned before, one obtains the imaginary part of the action,
\begin{eqnarray}
    Im \,\, S =  \int_m^\omega \frac{2 \pi}{f'(r_h)} d\omega', 
    \label{eqim}
\end{eqnarray}
where the radius of the horizon is $r_{h}$ is an explicit function of $(M - \omega^{'})$, where $M$ is the mass of the black-hole and $\omega^{'}$ is the energy of the tunneling particle under consideration. Now we make the following expansion for $f^{'}(r)$ and keep only the leading order term $f'(r_h(M))$ in the expansion,
\begin{eqnarray}
    f'(r_h) = f'(r_h)\Big{|}_{\omega^{'}= 0}-f''(r_h) \, \frac{\partial r_h}{\partial M}\Big{|}_{\omega^{'}= 0}  \omega^{'}  + \dots ~.
\end{eqnarray}
Following the steps described in \cite{article19}, we calculate the integral in Eq.(\ref{eqim}) using the above expression of $f'(r_h)$,
\begin{eqnarray}
    Im \,\, S = \frac{2 \pi}{f'(r_h)} (\omega - m) ~.
\end{eqnarray}
For $J \neq 0$ case, we have two horizons,
\begin{eqnarray}
    r_h^2 = r_{\pm}^2 = \frac{M}{2 \Lambda} \left[1 \pm \left(1 - \frac{\Lambda J^2}{M^2}\right)^{1/2}\right]~.
\end{eqnarray}
However, for our work, only the outer horizon is relevant. 
\begin{eqnarray}
    f'(r_h) = f'(r_+) = 2 \Lambda r_+ - \frac{J^2}{2r^3_+}~.
    \label{eqfdash}
\end{eqnarray}
For the outer horizon of the black-hole, the imaginary part of the action takes the following form,
\begin{eqnarray}
    Im \,\, S = \frac{\pi}{\sqrt{2 \Lambda}} \frac{\left(M + \sqrt{M^2 - \Lambda J^2}\right)^{1/2}}{\sqrt{M^2 - \Lambda J^2}} (\omega - m) ~,
\end{eqnarray}
and the corresponding expression of the tunneling amplitude is given by
\begin{eqnarray}
    \Gamma  \displaystyle \sim e^{-2 Im \,\, S} = e^{\frac{-2 \pi}{\sqrt{2 \Lambda}} \frac{\left(M + \sqrt{M^2 - \Lambda J^2}\right)^{1/2}}{\sqrt{M^2 - \Lambda J^2}}(\omega - m)} ~.
\end{eqnarray}
The expression of the tunneling amplitude clearly indicates that Hawking temperature \cite{article33, article34} ,
\begin{eqnarray}
    T_H = \frac{\sqrt{2 \Lambda}}{2 \pi} \frac{\sqrt{M^2 - \Lambda J^2}}{(M + \sqrt{M^2 - \Lambda J^2})^{1/2}}~.
\end{eqnarray}
Finally, following the steps described earlier for $J=0$ case, we get the expression of anyonic Hawking radiation spectrum for $J \neq 0$ case as
\begin{eqnarray}
    n(\omega, m) = \frac{\Gamma_{\omega}}{e^{\frac{2 \pi}{\sqrt{2 \Lambda}}\frac{(M + \sqrt{M^2 - \Lambda J^2})^{1/2}}{\sqrt{M^2 - \Lambda J^2}}(\omega - m)} - a(\alpha)} ~.
\end{eqnarray}
\section{Applications}
\label{exptsection}
\subsection{Experimental set-up }
In this section we review an analogue model of gravity as a potential system for testing our results. In these models, the dynamical equation of the analog system
closely resembles that of a quantum field in the background of a curved spacetime. This allows one to potentially test certain semi-classical and quantum gravitational results, especially those pertaining to Hawking radiation (see \cite{article38,article39,article32,article25,article24,article23} and the references therein). 
In particular, if one considers a two-dimensional photon superfluid system, it can be shown that the dynamics is governed by the equation of a massless scalar field in the background of an acoustic metric \cite{article40}. Furthermore, augmenting this equation by some corrections (last three terms in the RHS of Eq.(\ref{eqn60})), which may be realizable in the laboratory, we show that it governs the dynamics of anyons in the background of the superfluid or similar analog systems. 
It should be noted that Eq.(\ref{eqn60})) is nothing but the non-linear Schr\"odinger equation plus corrections,
\begin{eqnarray}
    \partial_{z} \Psi = \frac{i}{2 k} {\Bar{\nabla}}^{2} \Psi - \frac{ i k n_{2}}{n_{0}} \Psi \left|\Psi \right|^{2} + \frac{c\alpha}{n_{0}\Psi^{*}}  \partial_{z}\phi \nonumber\\
     + \frac{\alpha}{2 \Psi^{*}} ( \Bar{\nabla} \phi)^{2} + \frac{\beta}{\Psi^{*}} \phi~.    
     \label{eqn60}
\end{eqnarray}
Here, $z$ is the propagation direction and plays the role time. $c$ is the speed of light, $k$ is the wave number, $n_{2}$ is the material nonlinear coefficient and $n_{0}$ is the linear refractive index. $\Psi$ is the slowly varying envelope of electric field. So, $\mid\Psi\mid^{2}$ can be interpreted as the intensity of the optical field.
\begin{eqnarray}
\Psi \equiv \rho^{\frac{1}{2}} e^{i \phi}
~~~\text{and} \hspace{20pt}
    t= \frac{n_{0}}{c} z~.
    \label{eqn61}
\end{eqnarray}
The gradient operator `$\Bar{\nabla}$' is defined with respect to the transverse directions$(x,y)$ and, $\alpha$ and $\beta$ are real valued functions. The first two terms on the right hand side of Eq.(\ref{eqn60}) can be realized as the dynamics of massless scalar field in acoustic metric and the rest of the three terms are relevant for anyons. However, following a set of well-motivated assumptions, it is straight forward to show that the presence of the third term on the right hand side is sufficient to realize anyonic Hawking radiation.
Plugging Eq.(\ref{eqn61}) in Eq.(\ref{eqn60}) and separating the resultant equation in to real and imaginary part gives rise to the following equations,
\begin{equation}
    \partial_{t} \rho + \Bar{\nabla} . (\rho v) - 2 k \alpha \partial_{t} \psi - k \alpha v^{2} - 2 k \beta \psi =0~, \label{nlse1}
\end{equation}
\begin{equation}
    \partial_{t}\psi + \frac{1}{2} v^{2} + \frac{c^{2} n_{2}}{n_{0}^{3}} \rho = 0~, \label{nlse2}
\end{equation}
where
\begin{equation}
    v \equiv \frac{c}{k n_{0}} \Bar{\nabla} \phi 
    \equiv \Bar{\nabla} \psi~.
\end{equation}
Here $v$ can be interpreted as the fluid velocity. In the last step, following the articles \cite{article40,article22}, we have neglected the quantum pressure term which has no analogy in classical fluid dynamics.
Next, we linearize Eq.(\ref{nlse1}) and 
Eq.(\ref{nlse2}) around the background state ($\rho_{0}$,$\psi_{0}$) to obtain acoustic disturbances as first order fluctuations of quantities describing mean fluid flow,
\begin{equation}
    \rho = \rho_{0} + \epsilon \rho_{1} + O(\epsilon^{2} )~,
\end{equation}
\begin{equation}
    \psi = \psi_{0} + \epsilon \psi_{1} + O(\epsilon^{2})~,
\end{equation}
\begin{equation}
v_0 = \Bar{\nabla} \psi_0 ~.
\end{equation}
In terms of perturbed quantities, Eq.(\ref{nlse1}) and Eq.(\ref{nlse2}) take the following form in polar coordinates,
\begin{equation*}
    \partial_{t} \rho_{1} + \Bar{\nabla} . (\rho_{0} \Bar{\nabla} \psi_{1} + \rho_{1} v_{0}) - 2k\alpha \partial_{t} \psi_{1} - 2k\alpha v_{r} \partial_{r} \psi_{1}
\end{equation*}
\begin{equation}
    -2 k \alpha \frac{v_{\theta}}{r} \partial_{\theta} \psi_{1} - 2k\beta \psi_{1} = 0~,
    \label{eq67}
\end{equation}
\begin{equation}
    \partial_{t} \psi_{1} + v_{0} . \Bar{\nabla} \psi_{1} + \frac{c^{2} n_{2}}{n_{0}^{3}} \rho_{1} = 0~.
    \label{eq68}
\end{equation}
Eliminating $\rho_{1}$ from Eq.(\ref{eq67}) and Eq.(\ref{eq68}),
we obtain
\begin{eqnarray}
    &-&\partial_{t} \left(\frac{\rho_{0} }{c_{s}^{2}} \chi\right) + \Bar{\nabla} .\left(\rho_{0} \Bar{\nabla} \psi_{1} -\frac{\rho_{0}v_{0}}{c_{s}^{2}}\chi\right)  - 2 k \alpha \partial_{t} \psi_{1} \nonumber\\
    &-& 2 k \alpha v_{r} \partial_{r} \psi_{1}- 2k \alpha \frac{v_{\theta}}{r} \partial_{\theta} \psi_{1} - 2 k \beta \psi_{1} = 0~,
    \label{eq70}
\end{eqnarray}
where
\begin{eqnarray}
    \chi = \partial_{t} \psi_{1} + v_{0}.\Bar{\nabla} \psi_{1}  \quad \mathrm{and} \quad  c_{s} = \frac{c^{2}n_{2}\rho_{0}}{n_{0}^{3}}~.
\end{eqnarray}
Here $c_{s}$ is the local speed of sound. 
The final step is to set up a connection between dynamics of acoustic disturbances and the dynamics of an anyonic field, which can be written as an abelian Higgs model with a Chern-Simons term as follows \cite{article11},
\begin{eqnarray}
    g^{\mu \nu} \nabla_{\mu}\nabla_{\nu} \psi + 2 i q g^{\mu \nu} A_{\mu} \partial_{\nu} \psi &-& (2 c_{2} + q^{2} g^{\mu \nu} A_{\mu} A_{\nu} ) \psi \nonumber\\
     &+& 4 c_{4} \mid \psi \mid^{2} \psi = 0 ~.
    \label{acous1}
\end{eqnarray}
Here $c_{2}$ and $c_{4}$ are constants and $A_{\mu}$ is the four vector potential associated with the anyonic field. Consider Eq.(\ref{acous1}) in an acoustic metric with $c_{2}=c_{4}=0$ and 
$A_{\mu}= (i a_{} , 0, 0, 0)$ for an anyonic field $\psi_{1}$. Here we have considered an imaginary vector potential, similar to the assumption in \cite{article28}. With these choices of the parameters and the four vector potential, Eq.(\ref{acous1}) takes the following form,
\begin{eqnarray}
     g^{\mu \nu} \nabla_{\mu} \nabla_{\nu} \psi_{1} &+& \frac{2  q a_{}}{\rho_{0}^{2}} \partial_{t} \psi_{1} + \frac{2  q a_{} }{\rho_{0}^{2}}  v_{r} \partial_{r} \psi_{1}\nonumber\\
    &+& \frac{2 q a_{} }{\rho_{0}^{2}} \frac{v_{\theta}}{r}  \partial_{\theta} \psi_{1} + \frac{q^{2}}{\rho_{0}^{2}} a_{}^{2} \psi_{1} = 0 ~,
     \label{eq72}
\end{eqnarray}
and the `analog metric' is given by,
\begin{eqnarray}
    ds^{2} =\left( \frac{\rho_{0}}{c_{s}} \right)^{2} \left[ -  \left(1 - \frac{v^{2}}{c_{s}^{2}}\right) \, \left(c_{s}dt\right)^{2} - 2\frac{v_{r}}{c_{s}} \left(c_{s}dt\right) dr \right.\nonumber\\
    \left. - 2 \frac{v_{\theta}}{c_{s}} \left(c_{s}dt\right) \left(r\,d\theta\right) +  dr^{2} + \left(r\,d\theta\right)^{2}\right]~.
    \label{acmetric}
\end{eqnarray}
It is important to note that the overall structure of Eq.(\ref{eq70}) and Eq.(\ref{eq72}) are the same. This is the motivation for us to study anyonic Hawking radiation in the experimental set-up under consideration.  
Eq.(\ref{eq72}) is simplified further by assuming that the phase $\phi$ is slowly varying in space. This implies that $v_r$ and $v_\theta$ are very small. In addition to this we assume that the charge $q$ is also a negligible quantity. Under these assumptions, we can drop the last three terms in Eq.(\ref{eq72}),
\begin{eqnarray}
     g^{\mu \nu} \nabla_{\mu} \nabla_{\nu} \psi_{1} + \frac{2  q a_{}}{\rho_{0}^{2}} \partial_{t} \psi_{1} + O(\epsilon^{2})  = 0 ~.
     \label{eq74}
\end{eqnarray} 
Here we have taken $q \sim v_{i} \sim $ O$(\epsilon)$. 
It is now easy to see that the above equation is identical to Eq.(\ref{eq70}) under the previous assumptions.
\subsection{Hawking radiation of anyons from acoustic metric}
In section \ref{hranyonsection}, we have discussed about the methodology to calculate the tunneling amplitude of anyons from a BTZ black hole. In this section we will follow similar procedures to calculate the same in the case of acoustic metric (Eq.(\ref{acmetric})).
Just to be consistent with the metric of the geometry around BTZ black hole, we set $v_{\theta}=0$. In addition to this consider the case where $c_{s}$ is a constant.
In order to cast the Eq.(\ref{acmetric}) in the desired form we consider re-scaling of radial coordinates,
\begin{eqnarray}
    r \longrightarrow \frac{\rho_{0}}{c_{s}} r \qquad dr \longrightarrow \frac{\rho_{0}}{c_{s}} dr ~.
\end{eqnarray}
Under these re-scaling, the metric now takes a very simple form,
\begin{equation}
     ds^{2} = - \left( \frac{\rho_{0}}{c_{s}} \right)^{2} \left(1 - \frac{v_{r}^{2}}{c_{s}^{2}}\right)  (c_{s}dt)^{2} -   \frac{2 v_{r}\rho_{0}}{c_{s}^{2}}\,(c_{s}dt) dr +  dr^{2}~.
\end{equation}
This metric is further simplified by setting the radial component of the velocity vector in $c=1$ units as, $v_{r} = \displaystyle{-\frac{ \, \pi}{k n_{0} \sqrt{r r_{0}}}}$ \cite{article40,article22},
\begin{equation}
     ds^{2} = -f(r) dt^{2} +  2 f(r) \sqrt{\frac{1-g(r)}{f(r)g(r)}} dt dr +  dr^{2}~,
\end{equation}
where
\begin{equation}
    g(r) = 1 - \frac{r_{s}^{2}}{r_{0}\,r} \quad \mathrm{and} \quad f(r) =  \rho_{0}^{2}\,g(r) ~.
    \label{362}
\end{equation}
From the expression of $g(r)$, it is obvious that the horizon is located at $r = r_{h}$,
\begin{eqnarray}
    r_{h} = \frac{r^{2}_{s}}{r_{0}}~,
\end{eqnarray}
where
\begin{eqnarray}
r_{s} = \frac{\pi\,}{k\,n_{0}\,c_{s}} ~.
\end{eqnarray}
Given the current form of the metric, the imaginary part of the action is calculated as follows \cite{article19},
\begin{eqnarray}
         Im\,\,S  =  \int_{0}^{\omega} \, \frac{2\pi}{\sqrt{f'(\,r_{h}) \, g'(r_{h})}} \, d\omega' =  \frac{2 \, \pi\, r_{s}^{2}}{ \, \rho _0 \,r_0} \, \omega~,
\end{eqnarray}
where we have taken the lower limit as photon mass, $m$ to be zero.
It is now very straightforward to calculate the corresponding tunneling amplitude,
\begin{eqnarray}
    \Gamma  \displaystyle \sim e^{-2 \, Im\, \, S} \, = \displaystyle{e^{-\frac{\omega}{T_{H}}}} ~,
\end{eqnarray}
where
\begin{equation}
    T_{H }\, = \, \frac{r_0 \, \rho_0 \,}{4 \, \pi\, r_{s}^{2}}~
\end{equation}
is the Hawking temperature. 
Given this Hawking temperature, the distribution function of the particles in analogue model is given by,
\begin{equation}
    n(\omega) = \frac{\Gamma_{\omega}}{ e^{\frac{\omega}{T_{H}}} \, - a(\alpha)}~.
    \label{number1}
\end{equation}
We remind the reader that the parameter $a(\alpha)$ in Eq.(\ref{number1}) can take any
value in between $-1$ and $+1$ with $a(\alpha)=1$ and $a(\alpha)=-1$ corresponding to bosons and fermions respectively, and intermediate values signifying anyons. 
In principle the value of $a(\alpha)$ can be determined from the exact nature of anyons under consideration. However, if we are unaware about the statistics of the particle, experiment provides an alternate way to determine the value of parameter using Eq.(\ref{number1}). 
So, an experimental confirmation of the existence of probability distribution of anyons in a photon superfluid would not only support the existence of Hawking radiation, but also the confirm the existence of anyons. This work is the first attempt to demonstrate anyonic Hawking radiation to the best of our knowledge.
\section{Conclusion}
\label{conclusionsection}
In this work, we have derived a general expression for Hawking radiation of {\it anyons}, particles with intermediate statistics in $(2+1)$-dimensional spacetime, using the tunneling approach. The results are derived for both rotating and non-rotating BTZ black holes for massive and massless cases. We have shown that our results may be verifiable experimentally in an appropriate analogue system. Such a measurement on the one hand, will provide further evidence for Hawking radiation and that of anyons, albeit in an analog setting. Once anyon excitations are detected in this way, one can envisage potential uses of these particles with fractional statistics in a variety of ways. Furthermore, it is hoped that this work and its potential experimental verification would shed some light on the information loss problem. For example, our analysis is quantum mechanical and manifestly unitary, while the information loss problem suggests a fundamental non-unitarity. Thus it would be interesting to see how much of the current unitarity, say in analog models, can carry over to a real black hole and the subsequent Hawking radiation. 
We hope to report on these issues in the future. 
\section{Acknowledgement}
This work was supported by the Natural Sciences and
Engineering Research Council of Canada. 
\nocite{article1}\nocite{article2}\nocite{article3}\nocite{article5}\nocite{article6}\nocite{article7}\nocite{article8}\nocite{article9}\nocite{article11}\nocite{article12}\nocite{article13}\nocite{article15}\nocite{article16}\nocite{article18}\nocite{article19}\nocite{article20}\nocite{article21}\nocite{article22}\nocite{article23}\nocite{article24}\nocite{article25}\nocite{article28}\nocite{article29}\nocite{article30}\nocite{article31}\nocite{article32}\nocite{article33}\nocite{article34}\nocite{article35}\nocite{article36}\nocite{article37}\nocite{article40}\nocite{article41}\nocite{article43}\nocite{article44}
\bibliographystyle{ieeetr}
\bibliography{main}
\end{document}